\documentclass{scrartcl}
\usepackage{graphicx}
\usepackage{siunitx}
\usepackage{hyperref}
\begin{document}
\title{ The effect of wavefront aberrations
in atom interferometry}
\author{V. Schkolnik, B. Leykauf, M. Hauth, C. Freier, and A. Peters}                     
\date{}
\maketitle
\begin{center}\large Humboldt-Universit\"{a}t zu Berlin, Institut f\"{u}r Physik, \\
 Newtonstr. 15, 12489 Berlin, Germany \\
vladimir.schkolnik@physik.hu-berlin.de \\
\end{center}

\begin{center}
\textbf{Published in Applied Physics B\\
The final publication is available at Springer via\\
\href{http://dx.doi.org/10.1007/s00340-015-6138-5}{http://dx.doi.org/10.1007/s00340-015-6138-5}}
\end{center}

\begin{abstract}
Wavefront aberrations are one of the largest uncertainty factors in present atom interferometers. We present a detailed numerical and experimental analysis of this effect based on measured aberrations from optical windows. By placing windows into the Raman beam path of our atomic gravimeter, we verify for the first time the induced bias in very good agreement with theory. Our method can be used to reduce the uncertainty in atomic gravimeters by one order of magnitude, resulting in an error of less than \num{3e-10}\,$g$ and it is suitable in a wide variety of atom interferometers with thermal or ultra cold atoms. We discuss the limitations of our method, potential improvements and its role in future generation experiments.
\end{abstract}

\section{Introduction}
\label{intro}

In the last two decades, atom interferometry with cold atoms \cite{Kasevich1991} has become a versatile tool for high precision measurements of inertial forces like gravity \cite{Peters2001,Farah2014b,Hu2013,Hauth2013}, gravity gradients \cite{McGuirk2002}, and rotations \cite{Gustavson2000}. Moreover, atom interferometry is used for accurate measurements of the fine structure constant \cite{Wicht2002}, the gravitational constant \cite{Rosi2014}, tests of the weak equivalence principle \cite{Schlippert2014}, and could be used for gravitational wave detection \cite{Hogan2011}. Today, the sensitivity of gravity measurements with atom interferometry is higher than the best classical gravimeters that measure the gravitational acceleration with a falling corner cube \cite{Niebauer1995,Gillot2014} and are mainly limited by spurious vibrations.  Miniaturization of the atom interferometer components has led to mobile devices \cite{Hauth2013,Bidel2013,Gillot2014} that allow for measurements of the gravitational acceleration on sites of interest. 

The uncertainty of atom gravimeters, however, cannot yet compete with the uncertainty of the classical gravimeters \cite{Niebauer1995}. While effects like the Coriolis effect can be suppressed \cite{Lan2012,Hauth2014} and others like the AC Stark shift, magnetic field gradient and the two-photon light shift can be canceled by varying the effective Rabi frequency and alternating the orientation of the the beam splitter between a linear and a collinear orientation in respect to gravity \cite{Louchet-Chauvet2011a}, the effect of wavefront aberrations in these devices limit the uncertainty of atomic gravimeters to some \si{\micro Gal} (\SI{1}{\micro Gal} = \SI{e-8}{m/s^2} $\approx$ \num{e-9}\,$g$ is the traditional unit of gravity, named after Galileo Galilei). A recent publication \cite{Louchet-Chauvet2011a} reported the difficulty to measure the phase shift caused by wavefront aberrations in situ by extrapolating the temperature dependence of the phase shift to zero Kelvin. 

Our method is based on a detailed characterization of the wavefront aberrations introduced by optical elements, e.g. windows, prior to integration into the vacuum chamber of the apparatus. The effect of these aberrations on the atom interferometer phase can then be calculated taking into account the experimental parameters like atom cloud size, temperature, and the time between the interferometer pulses, among others.

The paper is structured as follows: Section 2 presents the Gravimetric Atom Interferometer GAIN realized at the Humboldt-University of Berlin and the sources of wavefront aberrations in this setup. In Section 3, we show the wavefront aberrations caused by high quality optical elements and a detailed study of their influence on the atom interferometer phase for different experimental parameters. In Section 4, we compare the measured  phase shift caused by an optical window inserted in the path of the Raman beam splitter with the results of our calculations. Finally, we discuss the limitations of the presented method, show possible improvements and emphasize the importance of a full understanding of wavefront aberrations for the next generation of atom interferometers.

\section{The Gravimetric Atom Interferometer GAIN}
\label{GAIN}

The mobile atom interferometer GAIN is a gravimeter based on a fountain of cold $^{87}$Rb atoms. A detailed overview of the experimental setup and the measurement sequence can be found in \cite{Schmidt2010} and \cite{Hauth2013}. 

\begin{figure}
\centering
  \includegraphics[width=0.75\textwidth]{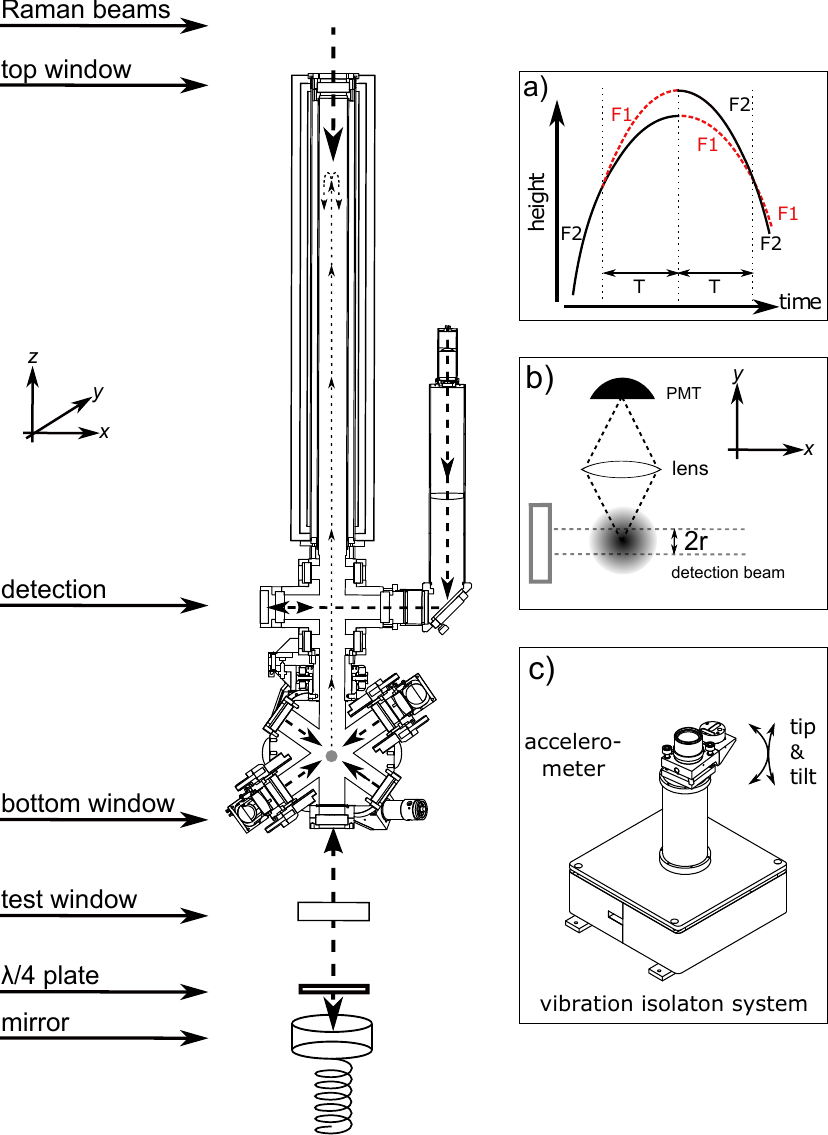}
\caption{A schematic overview of GAIN's vacuum chamber. Atoms are cooled and launched upwards in the MOT chamber. a) Three Raman pulses separated by $T$ = \SI{260}{ms} form a Mach-Zehnder interferometer. b) The  population of the atoms in the F=2 state is determined by normalized fluorescence detection. Photons emitted by the atoms in a standing wave are collected on a photomultiplier tube. c) The mirror for the Raman beams is mounted on a Tip/Tilt table on top of an active vibration isolator in order to align the retro-reflected beam.}
\label{fig:chamber} 
\end{figure}

About $10^8$ $^{87}$Rb  atoms are trapped in a Magneto-Optical Trap (MOT), launched upwards using the moving molasses technique \cite{Berthoud1999} and cooled adiabatically to about \SI{3}{\micro K} \cite{Drewsen1994}. After the launch, the atomic sample undergoes a selection and preparation sequence and enters a magnetically shielded interferometer zone in the  F=2, $m_{\mathrm{F}}$=0 hyperfine ground state. A narrow velocity distribution in vertical direction is selected by applying a long low-intensity Gaussian shaped Raman pulse \cite{Kasevich1991b}. The interferometer is realized by three Raman pulses in a $\pi/2$ -- $\pi$ -- $\pi/2$ configuration separated by  $T$ = \SI{260}{ms} (Figure~\ref{fig:chamber}a). The Raman beams enter the vacuum chamber through a polarization maintaining optical fiber from the top, leave the chamber through a bottom window and are retro-reflected by a mirror after passing a quarter-waveplate (see Figure~\ref{fig:chamber}). After the last pulse, the relative population of the atoms in the F=2 state is determined by normalized fluorescence detection. The transition probability 

\begin{equation}
 P_{| \mathrm{F}=2  \rangle } = \frac{1}{2} \left( 1 + \cos \Delta \Phi\right)
\label{population}
\end{equation}
depends on the phase of the Raman beams $\phi(t_i)$ at the time of the three pulses:

\begin{equation}
 \Delta \Phi = \phi(t_1) - 2 \cdot \phi(t_2) + \phi(t_3),
\label{phase}
\end{equation}
where $t_i$ are the times of the three Raman pulses with $t_2 = t_1 + T$ and $t_3 = t_2 + T$. Equation \ref{phase} assumes that the Raman beams are planar waves, so that all atoms experience the same phase shift $\phi(t_i)$ independent of their positions within the plane perpendicular to the beam direction. For arbitrary wavefronts, the phase shift for an atom $j$ depends on its position $\vec{r}_j$ within the beam profile and can be expressed as

\begin{equation}
 \Delta \Phi_j = \phi(\vec{r}_j(t_1)) - 2 \cdot \phi(\vec{r}_j(t_2)) + \phi(\vec{r}_j(t_3)),
\label{phase_r}
\end{equation}
As a result, the induced phase shift is not equal for all atoms within the atomic cloud. Therefore, the resulting phase error depends on the initial density distribution of the atomic cloud after the optical molasses, its expansion during the interferometer sequence and the averaging over the final spatial distribution during the detection process. Wavefront aberrations present in both counter-propagating Raman beams will cancel and only aberrations present in one of the beams contribute to the phase error in Equation~\ref{phase_r}. The effective wavefront error $\delta \phi_{\text{eff}}$ therefore can be expressed as 

\begin{equation}
 \delta \phi_{\text{eff}} = 2\cdot\delta \phi_{\text{w}} + 2\cdot\delta \phi_{\lambda/4} + \delta \phi_{\text{m}} ,
\label{wavefront_effective}
\end{equation}
where $\delta \phi_{\text{w}}, \delta \phi_{\lambda/4}$ and $\delta \phi_{\text{m}}$ are the aberrations from the bottom window, the quarter-waveplate, and the retro-reflecting mirror, respectively. In our setup the Raman beams have a diameter (1/$\mathrm{e}^2$) of \SI{29,5}{mm} and the corresponding Rayleigh length is about \SI{1}{km} so that changes of the wavefront due to beam divergence are neglected.

Mounting the optics introduces stress and deforms the wavefront. To successfully apply the formalism presented above, the wavefront measurements should be performed in the final mounting of the particular optical element or the influence of the mounting has to be known or measured. For optics placed outside the vacuum chamber (the quarter-waveplate and retro-reflecting mirror in our setup), this can be done easily.

For the bottom window, however, this is a more challenging task. Gluing or sealing the window with a wire made from an indium alloy, two vacuum sealing techniques widely used for hiqh quality windows in vacuum chambers, can introduce additional deformations in the order of the deformation of the unmounted element itself. Furthermore, measurements in a test setup showed that the pressure difference caused by evacuating the vacuum chamber mainly adds a defocus aberration of about $\lambda/10$ over an area with a diameter of \SI{20}{mm} on top of the arbitrary wavefront aberrations of the top and bottom window. One can measure the aberrations of this window by placing a telescope inside the atom interferometer's chamber or in a vacuum chamber with the same conditions of stress, connected to a light source via a fiber feed through and measure the wavefront in situ. This is of course only possible when the vacuum chamber is opened. For mobile devices, the change of the wavefront due to temperature changes or stress during transportation should be investigated. 

The error caused by wavefront aberrations in a cold atom gravimeter depends on numerous experimental parameters. We present the measurement of wavefront aberrations and the calculated error for the parameters used in our setup in the next section.

\section{Wavefront aberrations: measurement and calculation }
\label{Wavefront}

The wavefront aberrations for a set of 6 high quality windows are measured with a Shack-Hartmann sensor [SHSCam XHR GE, Optocraft]. In a differential measurement, the wavefront of a large collimated beam is subtracted as a reference wavefront from the wavefront transmitted through an optical window. Assuming no deformation of the sensor's micro lens array between the two measurements, only the wavefront errors of the window remain. The deformation is greatly reduced by temperature stabilization of the lens array by means of an integrated peltier element. The accuracy of this referenced method is $\sim\lambda/200$ rms over a circular area of  \SI{24}{mm} in diameter. 
The same method can be used for the quarter-waveplate or any other optical element working in transmission.

To obtain the wavefront error caused by the retro-reflecting mirror, a Shack-Hartmann sensor can be only used if a reference mirror with a known surface profile over the area of interest is available and the accuracy of the surface meets the needed requirements. An alternative approach is the use of optical interferometers that scan the mirror surface [e.g. NewView 6300, Zygo], thus requiring only a small reference flat, which is available more easily.   

\begin{figure}[htb]
\centering
\vspace*{5mm}
  \includegraphics[width=1.0\textwidth]{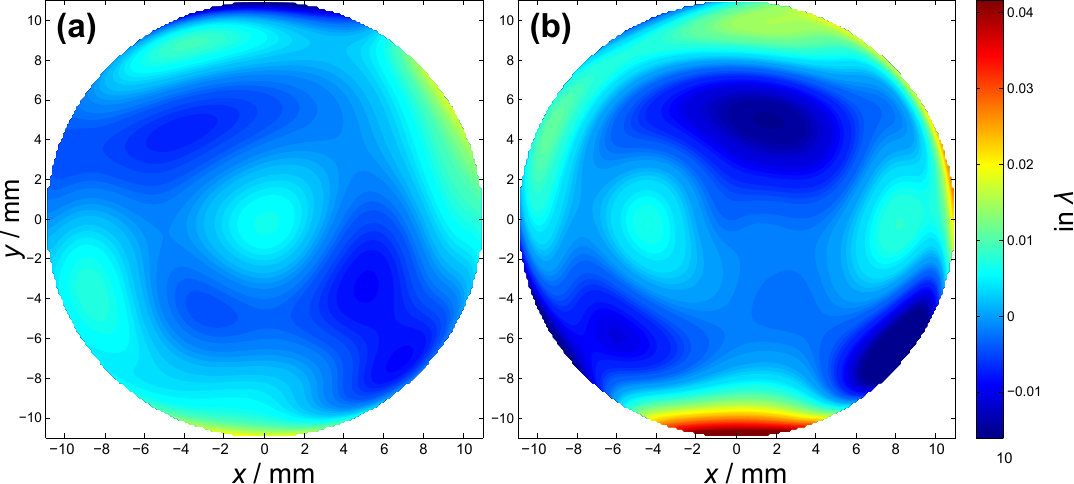}
\caption{The measured wavefront aberration from window no.\,1 (a) and window no.\,2 (b), represented by the first 36 Zernike polynomials. The piston and tilts are removed as they are actively canceled in our setup during the measurement.}
\label{fig:test1} 
\end{figure}

Independent of the method used, the wavefront is decomposed into Zernike polynomials and the first 36 coefficients are used for the following calculations. Figure~\ref{fig:test1} shows the reconstructed wavefront aberration of two windows.

\begin{figure}[htb]
\centering
  \includegraphics[width=0.7\textwidth]{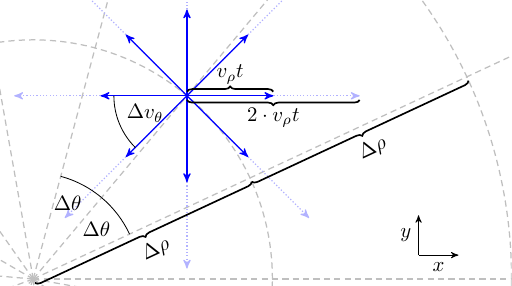}
\caption{Illustration of the model of the atomic cloud used in the numerical calculation. Test atoms are placed in the $x$-$y$ plane perpendicular to the Raman beams ($z$ direction) on each intersection of the dashed lines spaced by $\Delta \rho$ and $\Delta \theta$ in polar coordinates. Neglecting any interaction, the atoms' classical trajectories are straight lines. The arrows indicate position shifts after a time $t$ for two speeds $v_\rho$ and $2 \cdot v_\rho$ with isotropically distributed angles $\Delta v_\theta$.}
\label{fig:coordinates} 
\end{figure}

To calculate the phase shift caused by a window in our interferometer, we evaluate Equation \ref{phase_r} for our experimental parameters $t_1 = \SI{130}{ms}$ and $T = \SI{260}{ms}$, where $t_1$ is the time between launching the atomic cloud and the first Raman pulse and $T$ the time between the pulses. \SI{128}{ms} after the last Raman pulse, the atoms reach the detection zone and the fraction of the atoms in the exited state is determined. The detection light is delivered through a fiber. A lens with focal length of \SI{250}{mm} is used to collimate the detection beam to an 1/$\mathrm{e}^2$ diameter of \SI{50}{mm}, which gives a large central area with almost uniform intensity. The actual radius of the detection beam is set by an iris diaphragm positioned between fiber and lens. As a simplification of our detection model and because the exact response of our detection is not known yet, we assume that atoms inside a circular area with a radius $r$ contribute uniformly to the signal. The radius $r$ (see $r$ in Figure~\ref{fig:chamber}b) results in an intensity distribution close to a flat top beam for values of $r$ up to \SI{12}{mm}. 

Since our detection is not capable of a spatial resolution of the atomic cloud, we assume a two-dimensional Gaussian distribution of the atomic cloud in the horizontal plane with standard deviation $r_{\text{c}}$ = \SI{3}{mm} immediately after the launch. The radius $r_{\text{c}}$ of the atomic cloud after the launch was measured with a CCD camera. We also assume a two-dimensional Gaussian velocity distribution with a standard deviation $v_{\text{c}}$. This assumption is based on measurements of the velocity distribution of the atomic cloud in the $z$ direction using velocity selective Raman pulses. Although others found Lorentzian velocity distributions \cite{Farah2014}, we fitted the data to a Gaussian, as it yields the cloud temperature as a parameter. The differences in the simulation results between both distributions are negligible. The distributions are assumed to be in the plane of the effective wavefront. 

To deduce a value for the phase shift, this model was implemented in a numerical simulation. To this end, test atoms are placed in a plane perpendicular to the direction of beam propagation, as shown in  Figure~\ref{fig:coordinates}. Originating from these positions, the atoms' interaction-free, classical trajectories are calculated. For a given trajectory, the phase shift caused by wavefront aberrations is determined by evaluating Equation \ref{phase_r}. The total phase error is then determined by weighting these phase shifts by their respective spatial and velocity distributions and averaging all phase shifts contributing to the interferometer signal.

\begin{figure}[htb]
\centering
  \includegraphics[width=0.9\textwidth]{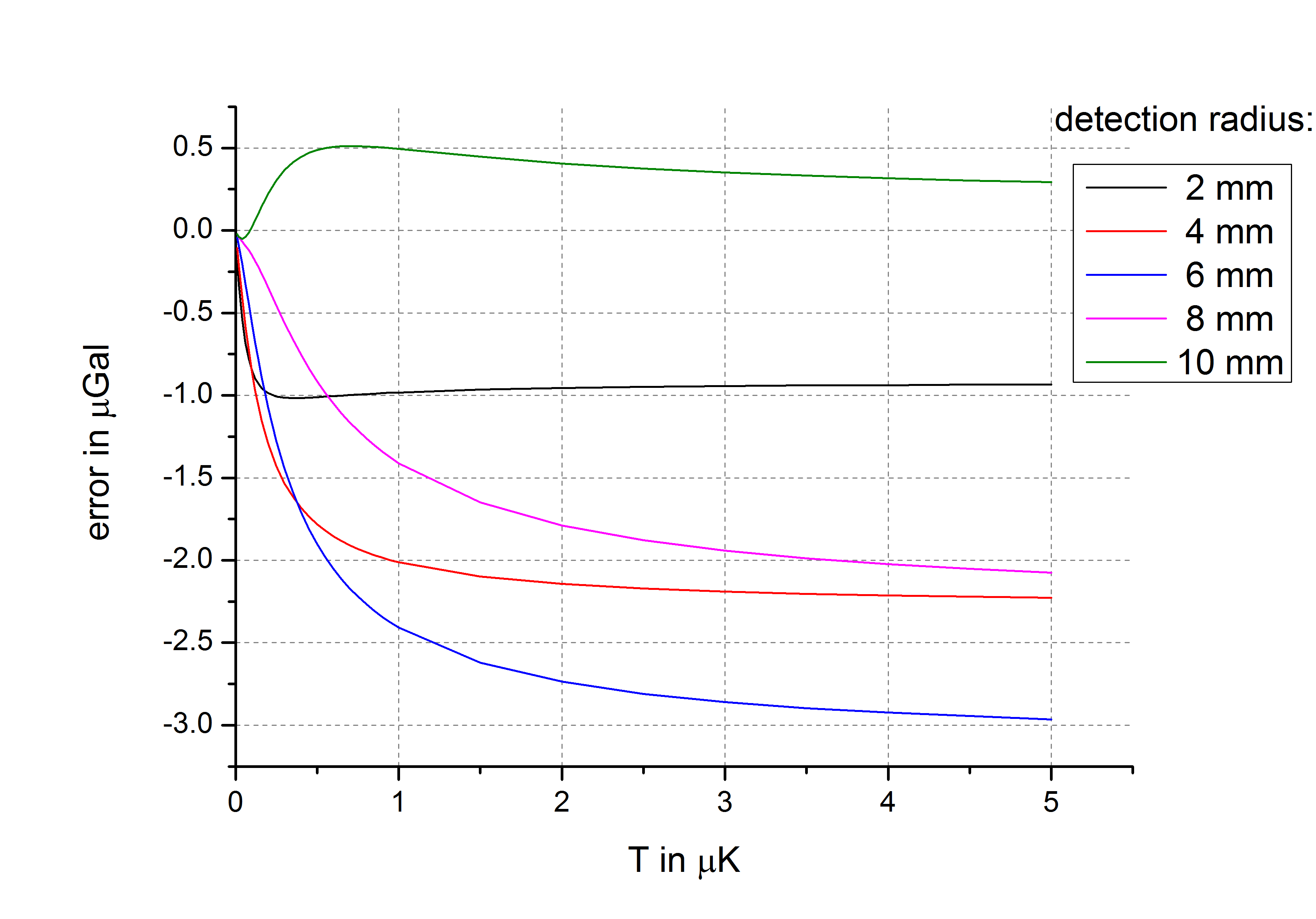}
\caption{The calculated temperature dependence of the wavefront induced error of window no.\,1 in our gravimeter for different detection radii.}
\label{fig:temperature_dependence} 
\end{figure}

Figure~\ref{fig:temperature_dependence} shows the calculated phase error in our gravimeter as a function of the cloud temperature for different detection radii. The spatial dependence of the wavefront error translates into a spatial dependence of the accumulated phase shift which is averaged over the detection area in our setup. The phase error as a function of the cloud's temperature shows only a minor change for higher temperatures, especially when reducing the temperature from \SI{3.5}{\micro K} (a temperature routinely achieved in our fountain) to about \SI{1.9}{\micro K} (the coldest temperature achieved in our setup). Colder temperatures are only achievable with Raman sideband cooling \cite{Kerman2000} or by evaporating the atomic cloud in a magnetic field or an optical dipole trap. The main reason for the weak temperature dependence lies in the fact that for temperatures corresponding to a final cloud size larger than the detection area, only atoms from the low kinetic energy tail remain in the detection zone and hotter atoms do not contribute to the signal at all. 

This is consistent with the result obtained by Louchet-Chauvet et al. \cite{Louchet-Chauvet2011a}, who tried to extrapolate the wavefront error to zero temperature, using a fit to Zernike polynomials. The main difficulty was the low sensitivity of the wavefront induced error to temperature changes in the regime achievable by optical molasses. This led to ambiguous results, depending on the order of the Zernike polynomials used for the fit. To reach an uncertainty below \SI{0.5}{\micro Gal} $\approx$ \num{5e-10 }\,$g$, temperatures less than \SI{200}{nK} are needed, more than one order of magnitude lower than achievable in our present setup.

Changing the temperature of the atomic cloud modifies the wavefront error only slightly. Changing the detection radius instead modifies the wavefront error over a wide range. In a recent study, Farah et al. \cite{Farah2014} demonstrated the importance of understanding the response of the detection in their atomic gravimeter by correcting the Coriolis bias (which is actively canceled in our setup) due to the convolution of the atomic cloud with the detection area. Fortunately, we are able to change the detection radius easily within our setup and are able to verify the calculated results and to correct the measured $g$ value. This can reduce the uncertainty of the error caused by this aberrations by a high degree without advanced cooling techniques. This is presented in the next section.

\section{Measurement of the phase shift caused by wavefront aberrations in an atom interferometer}

To show the potential accuracy of our method, we measure $g$ with and without the characterized window (window no.\,1) inserted into the beam path of the Raman beams (see Figure~\ref{fig:chamber}) for different detection radii. Since in our setup the retro-reflected beam is actively aligned co-linearly to better than \SI{1}{\micro rad} in respect to the incoming beam by back-coupling the retro-reflected beam into the fiber with a piezo driven tip-tilt mirror \cite{Hauth2013}, any tilt introduced by the window will be canceled. The Zernike coefficients corresponding to tip and tilt are thus set to zero for the following calculations.
For every data point we measure the gravitational acceleration $g_{\mathrm{W}}$ with our atom interferometer corrected for tides, air pressure and beam axis alignment with the window inserted, followed by a measurement of $g$ without the additional window. The induced error $\delta g$ is then the difference between both measurements $\delta  g = g_{\text{W}} - g$ with the standard deviation $ \sigma_{\delta} = \sqrt{\sigma_{\mathrm{W}}^2 + \sigma^2 }$, where $\sigma_{\mathrm{W}}$ and $\sigma$ are the standard deviations of the two measurements, respectively. Depending on the sensitivity due to the different detection radii, every single measurement is performed over a period of 5 to 25 hours. The detection radius is then changed and the process is repeated. 

\begin{figure}[htb]
\centering
  \includegraphics[width=1.0\textwidth]{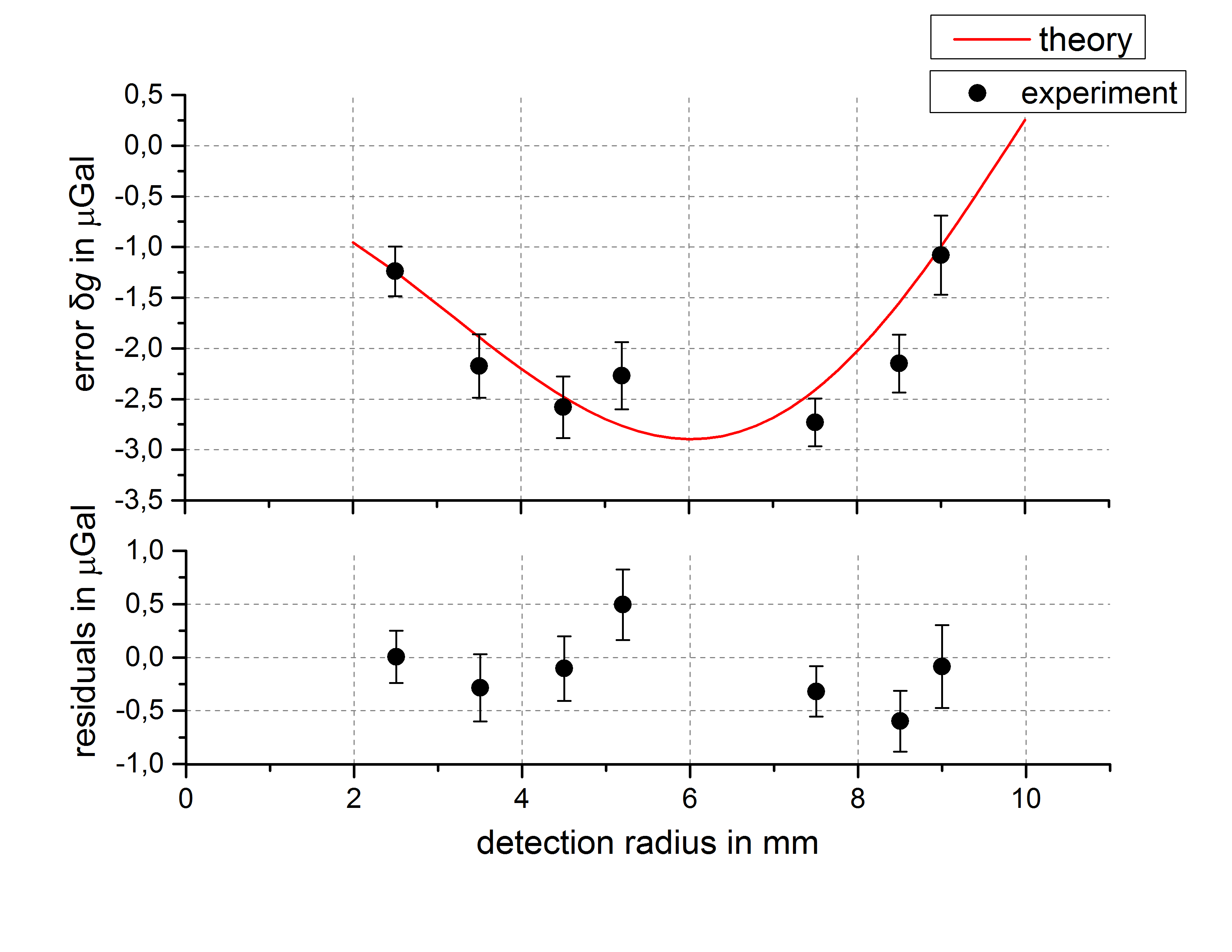}
\caption{Top: The measured error $\delta g$ of the gravitational acceleration caused by window no.\,1 inserted in the beam path of our gravimeter compared to the results obtained from theory. Bottom: The residuals show the difference between the measured and the calculated values. }
\label{fig:window1withresiduals} 
\end{figure}
The measured error in $g$ together with the calculated one is shown in Figure~\ref{fig:window1withresiduals}a. The measured data points are in excellent agreement with the theory. No fit of experimental parameters is performed and no offset is subtracted. Using this specific window in our setup and correcting the measured $g$ value for its aberrations results in a standard deviation of \SI{0.37}{\micro Gal} between the measured and the calculated value, almost a factor of 10 smaller than the effect itself. To our knowledge, this is the first time that the influence of wavefront aberrations is directly measured and compared successfully with theory. 

To show the utility of our method, we chose the window with the smallest calculated error from our batch, window no.\,2, shown in Figure~\ref{fig:test1}b and performed a second measurement with this window inserted into the beam path. The results are shown in Figure~\ref{fig:window1withresiduals}b. Again, an excellent agreement with calculated results is achieved with in a standard deviation of \SI{0.18}{\micro Gal} between measurement and theory, where the absolute error could already be reduced by a factor of 5 by selecting a window with wavefront aberrations resulting in a smaller error.

\begin{figure}[htb]
\centering
  \includegraphics[width=1.0\textwidth]{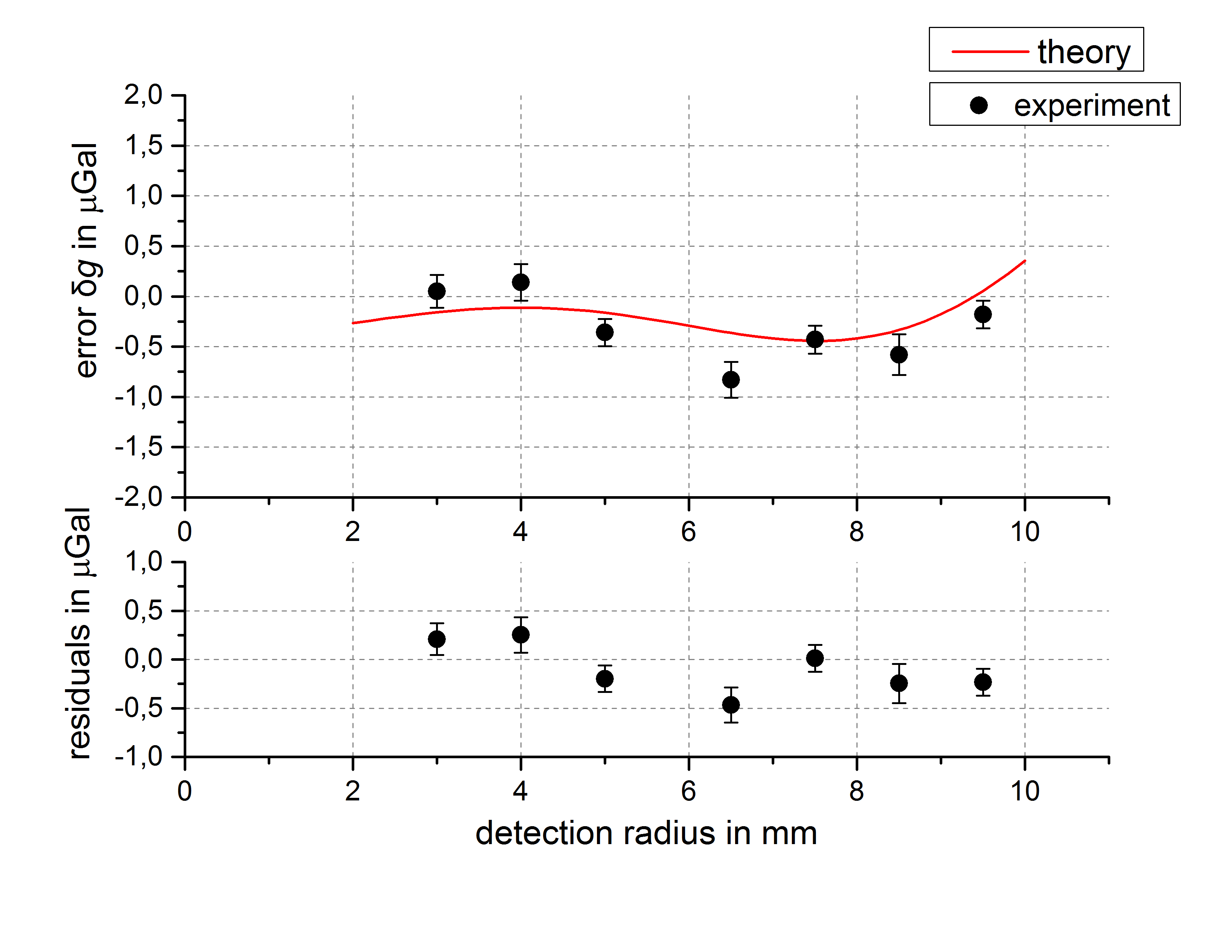}
\caption{The measured error for the window no.\,2 with the calculated results. The scale is the same as in Figure \ref{fig:window1withresiduals}. }
\label{fig:window2withresiduals} 
\end{figure}

Our method allows for characterization of optical elements prior to mounting and to select the ones with the smallest influence on the interferometers phase. The error can further be reduced by post-correction. Assuming the same achievable standard deviation for the quarter-waveplate and the mirror in our setup like in window no.\,2, the total uncertainty in our atom interferometer would be $\sqrt{9/4}$ $\cdot$ \SI{0.18}{\micro Gal} = \SI{0.27}{\micro Gal} or \num{2.7e-10}\,$g$, more than an order of magnitude smaller than the present uncertainty due to wavefront aberrations in cold atom gravimeters. Unfortunately, we are not able to measure the wavefront aberrations of our bottom window in situ at this point. A full characterization of all wavefront aberrations contributing to an offset in our gravimeter is in preparation.

\section{Conclusion and Outlook}
\label{sec:conclusion}

We demonstrated a method to correct for the offset caused by wavefront aberrations in an atomic gravimeter. The method is based on measuring the wavefront deformation caused by an optical element and calculating the induced error by numerical calculation. We demonstrated a reduction of the wavefront error (caused by a window) by a factor of 5 by selecting an appropriate window. Using our method for all optics in our setup could lead to an uncertainty of \num{2.7e-10}\,$g$ for the offset induced by wavefront aberrations. Without doubt, this approach can be widely used in present and future atom interferometers to reach a new level of accuracy within these devices.

While our method already produces remarkable results, it can further be improved. The implementation of a more elaborate detection mechanism able to resolve the interferometer phase spatially in the plane of the effective wavefront would improve the uncertainty considerably. A continuous monitoring of the size and position of the atomic cloud would correct for the error caused by long term drifts of these parameters. Our method requires measuring the wavefront before evacuating the vacuum chamber. Dickerson et al. \cite{Dickerson2013} could, however, establish a correlation between an atom's position and its velocity by using an initially point-like atomic source. Future progress of this method will allow for an in situ characterization of the effective wavefront.

Our method can also be used in differential atom interferometers where two or more atomic clouds share the same beam splitters. While the influence of wavefront aberrations in these devices is suppressed by the differential measurement, the effect of wavefront aberrations vanishes completely only when the atomic distribution of the clouds is exactly the same for the whole interferometer sequence. In a proposed test of the universality of free fall with a differential $^{87}$Rb/ $^{85}$Rb atom interferometer \cite{Schubert2013}, the requirements of the surface flatness for the optics used is very high. With a characterization of the optics mounted in the vacuum chamber these requirements can be relaxed.

\section*{Acknowledgments}

This work is supported by the European Science Foundation and the Deutsche Forschungsgemeinschaft (Euro\-Quasar-IQS, PE 904/2-1 and PE 904/4-1). We thank Andreas W. Schell for fruitful discussions and for helping with the data analysis.

\bibliographystyle{ieeetr}

\bibliography{Vladimir_Wavefront}

\begin{thebibliography}{10}

\bibitem{Kasevich1991}
M.~A. Kasevich and S.~Chu, ``{Atomic interferometry using stimulated Raman
  transitions},'' {\em Physical review letters}, vol.~67, no.~2, pp.~181--184,
  1991.

\bibitem{Peters2001}
A.~Peters, K.-Y. Chung, and S.~Chu, ``{High-precision gravity measurements
  using atom interferometry},'' {\em Metrologia}, vol.~38, p.~25, 2001.

\bibitem{Farah2014b}
T.~Farah, C.~Guerlin, A.~Landragin, P.~Bouyer, S.~Gaffet, F.~Pereira
  Dos~Santos, and S.~Merlet, ``Underground operation at best sensitivity of the
  mobile lne-syrte cold atom gravimeter,'' {\em Gyroscopy and Navigation},
  vol.~5, no.~4, pp.~266--274, 2014.

\bibitem{Hu2013}
Z.-K. Hu, B.-L. Sun, X.-C. Duan, M.-K. Zhou, L.-L. Chen, S.~Zhan, Q.-Z. Zhang,
  and J.~Luo, ``Demonstration of an ultrahigh-sensitivity atom-interferometry
  absolute gravimeter,'' {\em Phys. Rev. A}, vol.~88, p.~043610, Oct 2013.

\bibitem{Hauth2013}
M.~Hauth, C.~Freier, V.~Schkolnik, A.~Senger, M.~Schmidt, and A.~Peters,
  ``First gravity measurements using the mobile atom interferometer gain,''
  {\em Applied Physics B}, vol.~113, no.~1, pp.~49--55, 2013.

\bibitem{McGuirk2002}
J.~M. McGuirk, G.~T. Foster, J.~B. Fixler, M.~J. Snadden, and M.~A. Kasevich,
  ``Sensitive absolute-gravity gradiometry using atom interferometry,'' {\em
  Phys. Rev. A}, vol.~65, p.~033608, Feb 2002.

\bibitem{Gustavson2000}
T.~L. Gustavson, A.~Landragin, and M.~A. Kasevich, ``Rotation sensing with a
  dual atom-interferometer sagnac gyroscope,'' {\em Classical and Quantum
  Gravity}, vol.~17, no.~12, p.~2385, 2000.

\bibitem{Wicht2002}
A.~Wicht, J.~Hensley, E.~Sarajlic, and S.~Chu, ``{A preliminary measurement of
  the fine structure constant based on atom interferometry},'' {\em Physica
  Scripta}, vol.~2002, p.~82, 2002.

\bibitem{Rosi2014}
G.~Rosi, F.~Sorrentino, L.~Cacciapuoti, M.~Prevedelli, and G.~M. Tino,
  ``{Precision measurement of the Newtonian gravitational constant using cold
  atoms},'' {\em Nature}, vol.~510, pp.~518--521, June 2014.

\bibitem{Schlippert2014}
D.~Schlippert, J.~Hartwig, H.~Albers, L.~Richardson, L.\, C.~Schubert,
  A.~Roura, P.~Schleich, W.\, W.~Ertmer, and M.~Rasel, E.\, ``Quantum test of
  the universality of free fall,'' {\em Phys. Rev. Lett.}, vol.~112, p.~203002,
  May 2014.

\bibitem{Hogan2011}
J.~Hogan, D.~Johnson, S.~Dickerson, T.~Kovachy, A.~Sugarbaker, S.-w. Chiow,
  P.~Graham, M.~Kasevich, B.~Saif, S.~Rajendran, P.~Bouyer, B.~Seery,
  L.~Feinberg, and R.~Keski-Kuha, ``An atomic gravitational wave
  interferometric sensor in low earth orbit (agis-leo),'' {\em General
  Relativity and Gravitation}, vol.~43, no.~7, pp.~1953--2009, 2011.

\bibitem{Niebauer1995}
T.~M. Niebauer, G.~S. Sasagawa, J.~E. Faller, R.~Hilt, and F.~Klopping, ``A new
  generation of absolute gravimeters,'' {\em Metrologia}, vol.~32, no.~3,
  p.~159, 1995.

\bibitem{Gillot2014}
P.~Gillot, O.~Francis, A.~Landragin, F.~P.~D. Santos, and S.~Merlet,
  ``Stability comparison of two absolute gravimeters: optical versus atomic
  interferometers,'' {\em Metrologia}, vol.~51, no.~5, p.~L15, 2014.

\bibitem{Bidel2013}
Y.~Bidel, O.~Carraz, R.~Charrière, M.~Cadoret, N.~Zahzam, and A.~Bresson,
  ``Compact cold atom gravimeter for field applications,'' {\em Applied Physics
  Letters}, vol.~102, no.~14, pp.~--, 2013.

\bibitem{Lan2012}
S.-Y. Lan, P.-C. Kuan, B.~Estey, P.~Haslinger, and H.~M\"{u}ller, ``{Influence
  of the Coriolis Force in Atom Interferometry},'' {\em Physical Review
  Letters}, vol.~108, pp.~1--5, Feb. 2012.

\bibitem{Hauth2014}
M.~Hauth, C.~Freier, V.~Schkolnik, and A.~Peters, ``Atom interferometry for
  absolute measurements of local gravity,'' {\em Proceedings of the
  International School of Physics "Enrico Fermi"}, vol.~188, pp.~557--586,
  2014.

\bibitem{Louchet-Chauvet2011a}
A.~Louchet-Chauvet, T.~Farah, Q.~Bodart, A.~Clairon, A.~Landragin, S.~Merlet,
  and F.~P.~D. Santos, ``{The influence of transverse motion within an atomic
  gravimeter},'' {\em New Journal of Physics}, vol.~13, p.~065025, June 2011.

\bibitem{Schmidt2010}
M.~Schmidt, M.~Prevedelli, A.~Giorgini, G.~M. Tino, and A.~Peters, ``{A
  portable laser system for high-precision atom interferometry experiments},''
  {\em Applied Physics B}, vol.~102, pp.~11--18, Oct. 2010.

\bibitem{Berthoud1999}
P.~Berthoud, E.~Fretel, and P.~Thomann, ``Bright, slow, and continuous beam of
  laser-cooled cesium atoms,'' {\em Phys. Rev. A}, vol.~60, pp.~R4241--R4244,
  Dec 1999.

\bibitem{Drewsen1994}
M.~Drewsen, P.~Laurent, A.~Nadir, G.~Santarelli, A.~Clairon, Y.~Castin,
  D.~Grison, and C.~Salomon, ``Investigation of sub-doppler cooling effects in
  a cesium magneto-optical trap,'' {\em Applied Physics B}, vol.~59, no.~3,
  pp.~283--298, 1994.

\bibitem{Kasevich1991b}
M.~Kasevich, D.~S. Weiss, E.~Riis, K.~Moler, S.~Kasapi, and S.~Chu, ``Atomic
  velocity selection using stimulated raman transitions,'' {\em Phys. Rev.
  Lett.}, vol.~66, pp.~2297--2300, May 1991.

\bibitem{Farah2014}
T.~Farah, P.~Gillot, B.~Cheng, A.~Landragin, S.~Merlet, and F.~Pereira
  Dos~Santos, ``Effective velocity distribution in an atom gravimeter: Effect
  of the convolution with the response of the detection,'' {\em Phys. Rev. A},
  vol.~90, p.~023606, Aug 2014.

\bibitem{Kerman2000}
A.~J. Kerman, V.~Vuleti\ifmmode~\acute{c}\else \'{c}\fi{}, C.~Chin, and S.~Chu,
  ``Beyond optical molasses: 3d raman sideband cooling of atomic cesium to high
  phase-space density,'' {\em Phys. Rev. Lett.}, vol.~84, pp.~439--442, Jan
  2000.

\bibitem{Dickerson2013}
S.~M. Dickerson, J.~M. Hogan, A.~Sugarbaker, D.~M.~S. Johnson, and M.~A.
  Kasevich, ``Multiaxis inertial sensing with long-time point source atom
  interferometry,'' {\em Phys. Rev. Lett.}, vol.~111, p.~083001, Aug 2013.

\bibitem{Schubert2013}
C.~Schubert, J.~Hartwig, H.~Ahlers, K.~Posso-Trujillo, N.~Gaaloul, U.~Velte,
  A.~Landragin, A.~Bertoldi, B.~Battelier, P.~Bouyer, F.~Sorrentino, G.~M.
  Tino, M.~Krutzik, A.~Peters, S.~Herrmann, C.~L{\"a}mmerzahl, L.~Cacciapouti,
  E.~Rocco, K.~Bongs, W.~Ertmer, and E.~M. Rasel, ``{Differential atom
  interferometry with $^{87}$Rb and $^{85}$Rb for testing the UFF in
  STE-QUEST},'' {\em ArXiv e-prints}, Dec. 2013.

\end{thebibliography}

\end{document}